\documentclass[epj,final]{svjour}

\usepackage{graphicx}
\usepackage{dcolumn}   
\usepackage{bm}        
\usepackage{amssymb}   
\usepackage{textcomp}
\usepackage{ucs}
\usepackage{hyperref}
\usepackage{amsmath}
\usepackage{color}
\usepackage{footnote}

\begin{document}

\title{Intense Infrared Scintillation of Liquid Ar-Xe Mixtures}

\author{A. Neumeier\inst{1} \and T. Dandl\inst{2} \and T. Heindl\inst{2} \and A. Himpsl\inst{2} \and H. Hagn\inst{1} \and M. Hofmann\inst{1,4} \and L. Oberauer\inst{1} \and W. Potzel\inst{1} \and S. Roth\inst{1} \and \\S. Sch\"onert\inst{1} \and J. Wieser\inst{3} \and A. Ulrich\inst{2}\thanks{\emph{Andreas Ulrich:} andreas.ulrich@ph.tum.de 
}
}                   
\institute{Technische Universit\"at M\"unchen, Physik-Department E15, James-Franck-Str. 1, D-85748 Garching, Germany \and Technische Universit\"at M\"unchen, Physik-Department E12, James-Franck-Str. 1, D-85748 Garching, Germany \and excitech GmbH, Branterei 33, 26419 Schortens \and now at: KETEK GmbH, Hofer Str. 3, 81737 M\"unchen, Germany}

\date{Published in EPL (2014)}

\abstract{
Intense infrared (IR) light emission from liquid Ar-Xe mixtures has been observed using 12\,keV electron-beam excitation. The emission peaks at a wavelength of 1.18\,$\mu$m and the half-width of the emission band is 0.1\,$\mu$m. Maximum intensity has been found for a 10\,ppm xenon admixture in liquid argon. The conversion efficiency of electron beam-power to IR-light is about 1\,\% (10000 photons per MeV electron energy deposited). A possible application of this intense IR emission for a new  particle discrimination concept in liquid noble gas detectors is discussed. No light emission was found for perfectly purified liquid argon in the wavelength range from 0.5 to 3.5\,$\mu$m on the current level of sensitivity.
}         

\maketitle

\section{Observation}

\begin{figure}
  \centering
  \includegraphics[width=\columnwidth]{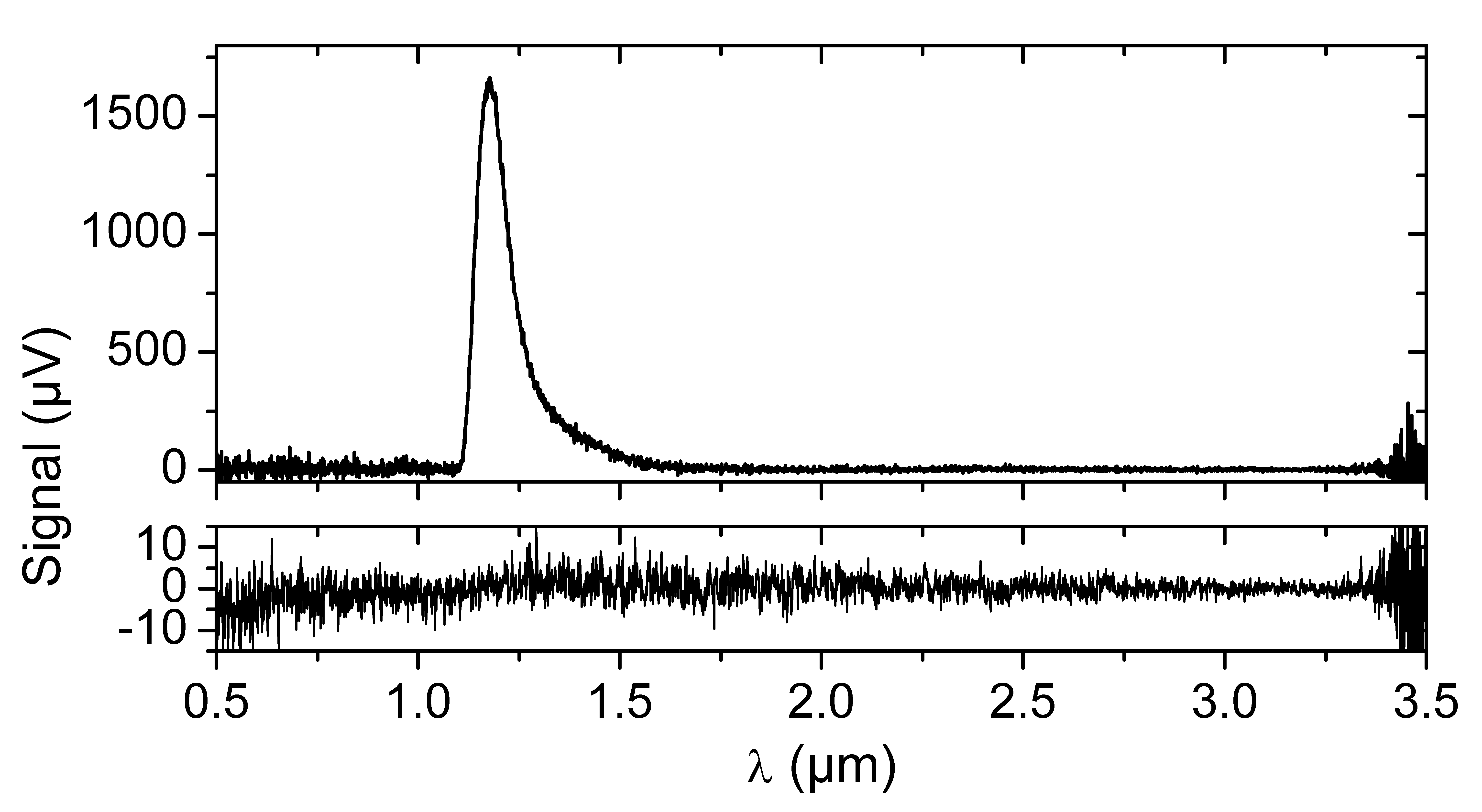}
    \caption{\textit{Electron-beam induced near-infrared emission spectrum of a mixture of 10 ppm xenon in liquid argon (upper panel). The broadband emission feature starts at $\sim$\,1.1\,$\mu$m, peaks at 1.18\,$\mu$m and ends at $\sim$\,1.6\,$\mu$m. The half-width of the asymmetrically shaped emission structure is 0.1\,$\mu$m. The measured IR emission of pure liquid argon is shown in the lower panel. It can be seen that a potential emission in pure liquid argon is at least a factor of 150 weaker than the emission from the mixture. The increased noise towards the long-wavelength end of the spectra is an artifact due to normalization by the detector response function.}}
 \label{fig:IR_Emission_LAr10ppmXe_LAr}
\end{figure}

We report that electron-beam excitation of liquid argon with a small admixture of xenon leads to a very intense emission band in the near-infrared spectral region. This emission dominates the infrared (IR) spectrum in a similar way as the excimer-like emission bands in liquid argon and liquid xenon {do} in the vacuum ultraviolet (VUV) spectral region \cite{Heindl_1}. The wavelength position, the spectral width, and the shape of the emission are shown in the spectrum of Fig.\,\ref{fig:IR_Emission_LAr10ppmXe_LAr} (upper panel). The {maximum} peak intensity lies at 1.18\,$\mu$m. The emission structure starts at 1.1 {(5\,\% of maximal intensity)} and ends at 1.5\,$\mu$m {(5\,\% of maximal intensity)} and has an asymmetric shape. The spectrum was recorded with an admixture of 10\,ppm xenon added to argon. This mixture was prepared in the gas phase prior to condensation. A 12\,keV (DC) electron-beam of 1.3\,$\mu$A beam current was used for exciting the liquid noble 
gas mixture. To the best of our knowledge{,} this {is the first observation of this emission}. So far 
we have no clear assignment to the species which is responsible for this IR emission. The shape and width of the emission hints towards an excimer-type transition \cite{Excimer_Buch} between a bound and a repulsive state in ArXe$^{**}$ or Xe$_{2}^{**}${\footnote{{Following the nomenclature in gas kinetics * denotes the first excited state and ** denote highly excited states of atoms and molecules, respectively.}}}. For comparison, the lower panel of Fig.\,\ref{fig:IR_Emission_LAr10ppmXe_LAr} shows the measured IR emission of pure liquid argon. No emission was found in this spectral range.

\section{Motivation}

The emission was found in the context of a series of experiments on the emission and absorption features of liquid noble gases \cite{Heindl_1,Heindl_2,Hofmann,Neumeier}. This study was motivated by the fact that liquid rare gases are presently often used in large quantities as the detector material in rare event physics, e.g., the search for the neutrinoless double-beta decay (GERDA Phase II uses a liquid argon veto \cite{GERDA}, EXO {uses liquid xenon} \cite{EXO}) and the direct dark matter search (Liquid argon: {WArP \cite{WARP},} ArDM \cite{ArDM_1,ArDM_2}, DarkSide \cite{DarkSide}; Liquid xenon: XENON100 \cite{XENON100}, LUX \cite{LUX}). {New experiments using liquid rare gas targets at the ton scale are currently under construction (XENON1T \cite{XENON1T}, MiniCLEAN \cite{MiniCLEAN}, DEAP-3600 \cite{DEAP}) and multi-ton projects (DARWIN \cite{DARWIN_1,DARWIN_2}) are conceived.} In the sector of high-energy neutrino physics (ICARUS \cite{ICARUS}, GLACIER \cite{GLACIER}, LBNE \cite{LBNE}) there are also 
extensive research and development activities to use liquid rare gas detectors. The optical properties of liquid noble gases are a key parameter to fully understand the detector modules. This aspect should be emphasized, in particular, due to the increasing detector volumes.

{Claims about infrared emission from liquid argon have previously been made} \cite{Italiener_IR_1,Russen_IR_1,Russen_IR_2,Russen_IR_3}. A spectrum in a previous publication \cite{Heindl_1} also showed a hint for an increasing intensity in the near-infrared. This motivated us to monitor the wavelength region between 0.5 and 3.5\,$\mu$m. We have now reached a point in our experiments which allows us to give up the restriction to work only with argon to prevent contamination of pure argon with xenon which had been identified as a problematic impurity both in emission \cite{Heindl_1} and absorption \cite{Neumeier}. The possibility to remove xenon from argon very efficiently (see next section) allowed us to dope argon with xenon in a controlled way.

\section{Experimental Setup}

The experimental setup used here has been described in detail in ref. \cite{Heindl_2}. Briefly, a 12 keV electron-beam is sent through a thin (300\,nm) silicon nitride/silicon oxide membrane \cite{Wieser_Membran} to excite liquid noble gases. {Emitted light} is collected by reflective optics and sent into a f = 30\,cm scanning monochromator with an Al-MgF$_{2}$ coated grating with 300\,$\frac{\textrm{lines}}{\textrm{mm}}$ and 1\,$\mu$m blaze wavelength. The response function of the monochromator has not been measured. However, based on the monochromator specifications we assume that the shape of the emission {shown in Fig.\,\ref{fig:IR_Emission_LAr10ppmXe_LAr}} is a good approximation to the {actual} spectral shape of the emission. The spectra are normalized with the detector responsivity provided by the manufacturer of the detector. The wavelength resolution of the 
detection system with a slit width of 500\,$\mu$m is 0.0052\,$\mu$m. In the present study the spectrometer was equipped with a thermoelectrically cooled InAs IR detector (Teledyne Judson, 
Model: J12TE3-66D-R01M-AST). Its signal was amplified with a transimpedance pre-amplifier (Teledyne Judson, Model: PA-6-60) and sent to a lock-in amplifier. A chopper wheel was placed in front of the spectrometer to chop the light and to provide the reference signal for the lock-in amplifier. Every spectrum presented has carefully been corrected for background radiation (mainly thermal radiation of the glowing filament of the cathode ray tube).

A key experimental aspect is gas handling. The excited levels of noble gases lie at such a high energy that they readily transfer energy to any kind of impurities{,} which leads to energy loss for the light production. The gas is therefore constantly circulated through a rare gas purifier (SAES getters, Model: MONO TORR$^{\textrm{\textregistered}}$ PS4-MT3-R-2) to remove contaminations{,} such as water {released} from the walls{, and} residual oxygen and nitrogen. For the purification of argon from heavier noble gases a distillation procedure has been {added} to the setup described in ref.\,\cite{Heindl_2}. {New improvements} to the measurements presented in ref. \cite{Neumeier} {allow the} cell for distillation {to be} precisely temperature controlled and the gas is kept in a continuous flow through the distiller. This modification allows {the removal of} xenon from argon so efficiently that no spectroscopic features related to 
xenon are 
visible {either} in emission {or} absorption spectra. Very 
small quantities of xenon can now be added to argon in a controlled way. The mixtures were prepared in the gas phase using a precision membrane capacity gauge (MKS Baratron 390H 1000).

{It should be mentioned that the xenon concentration in the liquid argon can to some extent be increased by condensation out of the buffer volume. Approximately 40\,\% of the initial gas volume are condensed. This corresponds to a maximum deviation of the xenon concentration in the liquid phase by a factor of $\sim$\,2.5. On the other hand the xenon concentration in the liquid could be reduced due to condensation of xenon on the walls of the cryogenic cell.}

\section{Spectroscopic Results}

The emission spectrum shown in Fig. \ref{fig:IR_Emission_LAr10ppmXe_LAr} (upper panel) is part of a series of spectra recorded with various xenon concentrations in liquid argon. A first important result of the study is that pure argon (Fig.\,\ref{fig:IR_Emission_LAr10ppmXe_LAr}, lower panel) shows no emission in the spectral region from 0.5 to 3.5\,$\mu$m studied with the solid-state IR detector attached to the spectrometer. A potential emission feature in pure liquid argon is at least $\sim$150 times weaker than the emission in the 10\,ppm argon-xenon mixture. Thereby, we have to revise results of earlier experiments (Fig. 2 in ref. \cite{Heindl_1}) which showed an increase in intensity towards the IR region. We now believe that this was an artifact due to the normalization of the spectrum with the response function of the spectrometer used for that spectral region. Unfortunately, these data have meanwhile been reproduced several times in the literature \cite{Russen_IR_2,Russen_IR_3}.

\begin{figure}
  \centering
  \includegraphics[width=\columnwidth]{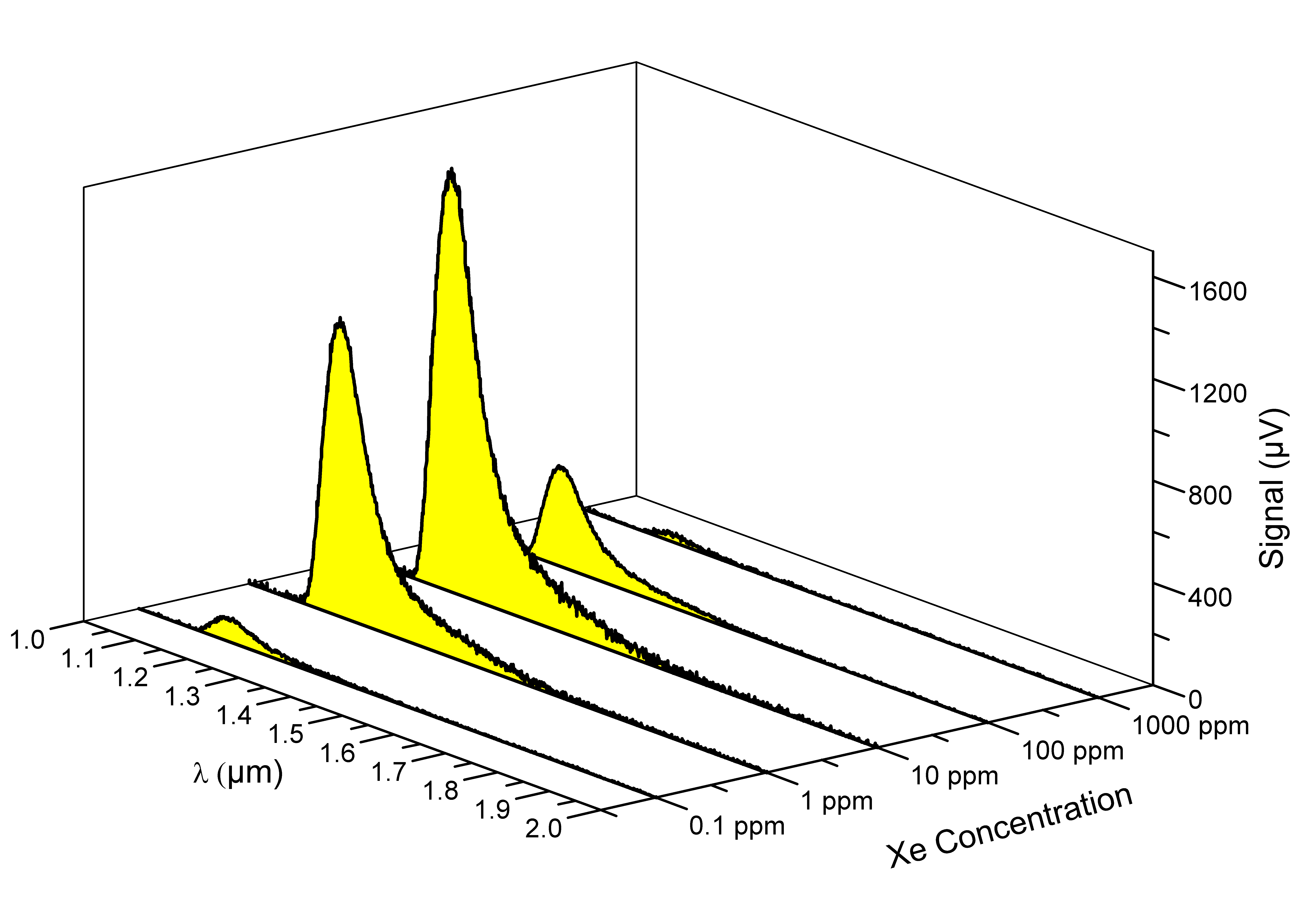}
  \caption{\textit{Electron-beam induced IR emission spectra of 0.1, 1, 10, 100 and 1000\,ppm xenon in liquid argon. The most intense IR emission was found for the mixture with 10 ppm xenon in liquid argon.}}
  \label{fig:3D_IR_Emission_LArXe}
\end{figure}

\begin{figure}
  \centering
  \includegraphics[width=\columnwidth]{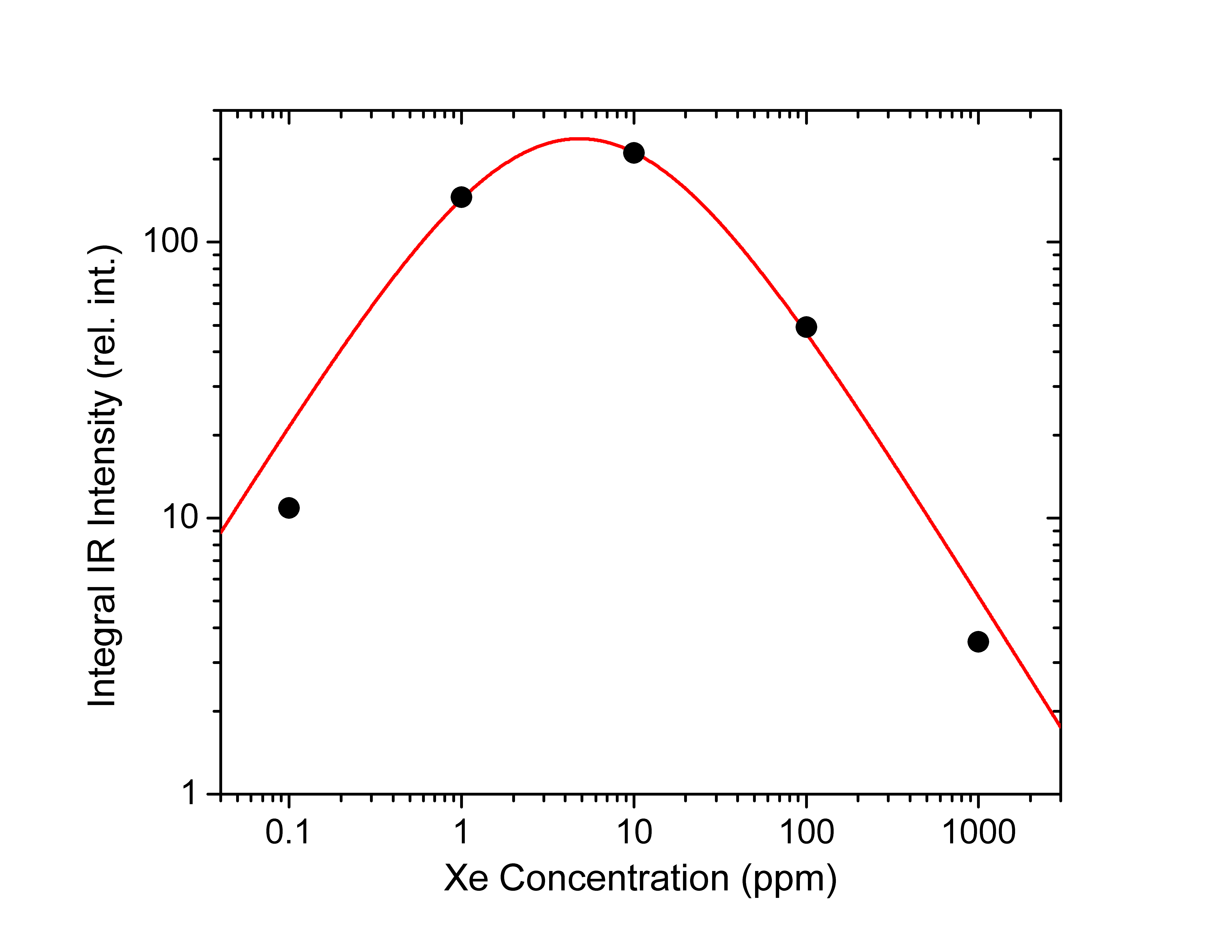}
  \caption{\textit{The plot shows the wavelength integrated IR emission of electron-beam induced scintillation of liquid argon with admixtures {of xenon at 5 different concentrations} (black dots) as obtained from the data in Fig. \ref{fig:3D_IR_Emission_LArXe}. The red solid curve shows a simple model calculation, described in the text. Note the double logarithmic scale and that the xenon concentration extends over four orders of magnitude.}}
  \label{fig:Integral_IR_Emission_LArXe}
  \end{figure}

An overview over a series of liquid argon-xenon spectra where xenon was added {at} 5 different concentrations to argon before condensation is shown in Fig. \ref{fig:3D_IR_Emission_LArXe}. It demonstrates that only a very small amount of 10\,ppm xenon is necessary to maximize the intensity of the newly found 1.18\,$\mu$m emission band. The different mixtures and the corresponding results could be reproduced very well ($\pm$\,10\%) in a series of experiments. A double logarithmic plot of the wavelength-integrated intensity of the emission versus xenon concentration is shown in Fig. \ref{fig:Integral_IR_Emission_LArXe} (black dots).

A detailed gas kinetic study using pulsed excitation and a time-resolved measurement of the light output will be necessary to determine the rate constants for the energy transfer from liquid argon to the xenon or argon-xenon emission band.


\begin{figure}
    \centering
    \includegraphics[width=\columnwidth]{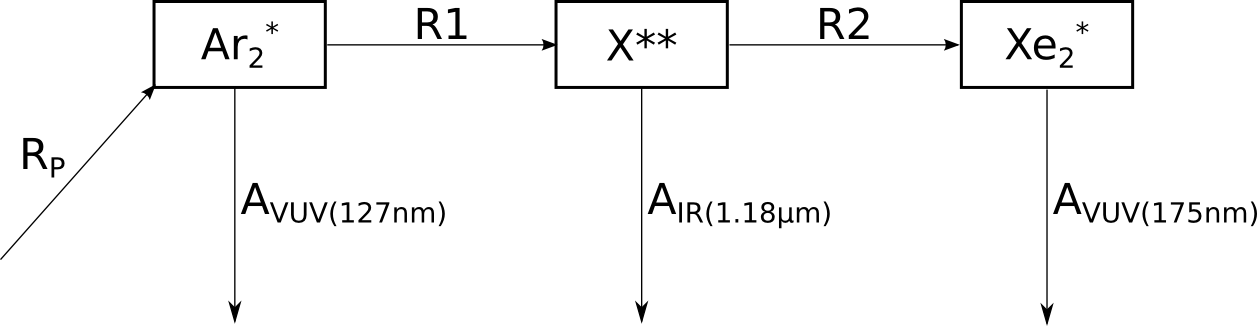}
    \caption{\textit{The flow diagram shows a simple kinetic model to explain the data shown in Fig.\,\ref{fig:Integral_IR_Emission_LArXe} (see text).}}
    \label{fig:Modell_LArXe_IR_Emission}
 \end{figure}

The trend of the data can, however, already be explained by a simple kinetic model which is visualized in Fig.\,\ref{fig:Modell_LArXe_IR_Emission}. Argon is excited at a rate R$_{\textrm{P}}$. Excimer-like states (Ar$_{2}^{*}$) form and decay emitting VUV light (A$_{\textrm{VUV}(127nm)}$). When xenon is added the IR emitting so far unknown species X$^{**}$ is formed at a rate R1. For the decay of the X$^{**}$-species there is a branching ratio between the IR light emission (A$_{\textrm{IR}(1.18\mu m)}$) and a reaction which transfers the energy at a rate R2 further on to xenon species for increasing xenon concentrations, most likely Xe$_{2}^{*}$,  which is already visible by its decay (A$_{\textrm{VUV}(175nm)}$) in the liquid argon spectra with xenon impurities in ref. \cite{Heindl_2}. Formulating the rate equations for the reactions in Fig.\,\ref{fig:Modell_LArXe_IR_Emission} and performing a fit to the data in Fig.\,\ref{fig:Integral_IR_Emission_LArXe} leads to the red (solid) line shown in Fig.\,\ref{fig:Integral_IR_Emission_LArXe}. Two-body collisions were assumed for the energy transfer processes. This shows that the simple model can reproduce the trend in the data. However, the error in the fit parameters is very large due to the limited number of data points spread over four orders of magnitude of xenon concentration. We can not yet quantify the values of the rate constants for the energy transfer without a measurement of the lifetime of species X$^{**}$ (A$_{\textrm{IR}(1.18\mu m)}$). An upper limit of the decay time (90\,\% to 10\,\% level) is 8.8\,${\mu}$s. This value is estimated from a wavelength-integrated measurement with a pulsed electron-beam and is dominated by the response of the IR detection system.

\section{Efficiency Measurement}

To judge whether the newly found infrared emission could be important for particle detector development we have measured the efficiency of the light output to {the best} precision which is presently possible using a setup which was slightly modified with respect to the spectroscopic studies. The InAs detector was removed from the monochromator and placed at a distance of 20\,cm from the beam-excited noble gas mixture. Consequently the IR emission could be measured with a well defined solid angle and all optical elements except for the MgF$_{2}$ exit windows were removed. The electron-beam was periodically switched on and off with a duty cycle of 50\% and a repetition rate of 5\,kHz using a pulsing feature available from the power supply of the electron gun. The signal was amplified by a factor of $10^{6}$ with the transimpedance pre-amplifier and AC-coupled to the lock-in amplifier for steering the beam into the cell by maximizing the signal. Note that AC coupling is necessary since the 
signal the detector receives from the warm environment is about 100 times larger than the signal from the small electron-beam excited, IR-emitting spot. For determining the absolute 
signal the detector output was sent to a storage oscilloscope directly from the pre-amplifier, digitally filtered (low-pass, -3\,dB at 80\,kHz) and averaged over 10000 traces. The result is that a 0.7\,$\mu$A, 7\,mW beam (average current, 1.4\,$\mu$A during the pulses) caused a detector signal corresponding to $10^{-10}$\,A. With the sensitivity of 0.5\,$\frac{\textrm{A}}{\textrm{W}}$ of the detector at 1.18\,$\mu$m wavelength and a solid angle of $1.6\cdot10^{-6}$ we find an energy efficiency of $\sim$1\,\% for the IR emission with electron-beam excitation. This converts into a value of $\sim$\,10000 IR-photons per MeV electron energy deposited. Assuming that there is no collective effect in the excitation process {by} a particle beam this value should also hold for scintillation light induced in a particle detector by single electrons. In the present experiments we found a linear behavior with beam current down to 0.2\,$\mu$A during the pulses. For comparison, the VUV-emission in pure 
gaseous argon excited by 10\,keV electrons has an efficiency of $(33 \pm 4)$\,\% \cite{Morozov_Effizienz}. The VUV intensity in the liquid phase was found to be 59\,\% of the intensity in the gas phase \cite{Heindl_2}. The liquid phase has a scintillation yield of $(1.9 \pm 0.3) \cdot 10^{4}$ VUV photons per MeV electron energy deposited \cite{Heindl_2}. {In the liquid} Doke et al. found a scintillation yield for 1\,MeV electron excitation of approximately 41000 photons per MeV deposited \cite{Doke_Effizienz_1,Doke_Effizienz_2}.

\section{Discussion}

\begin{figure}
 \centering
 \includegraphics[width=\columnwidth]{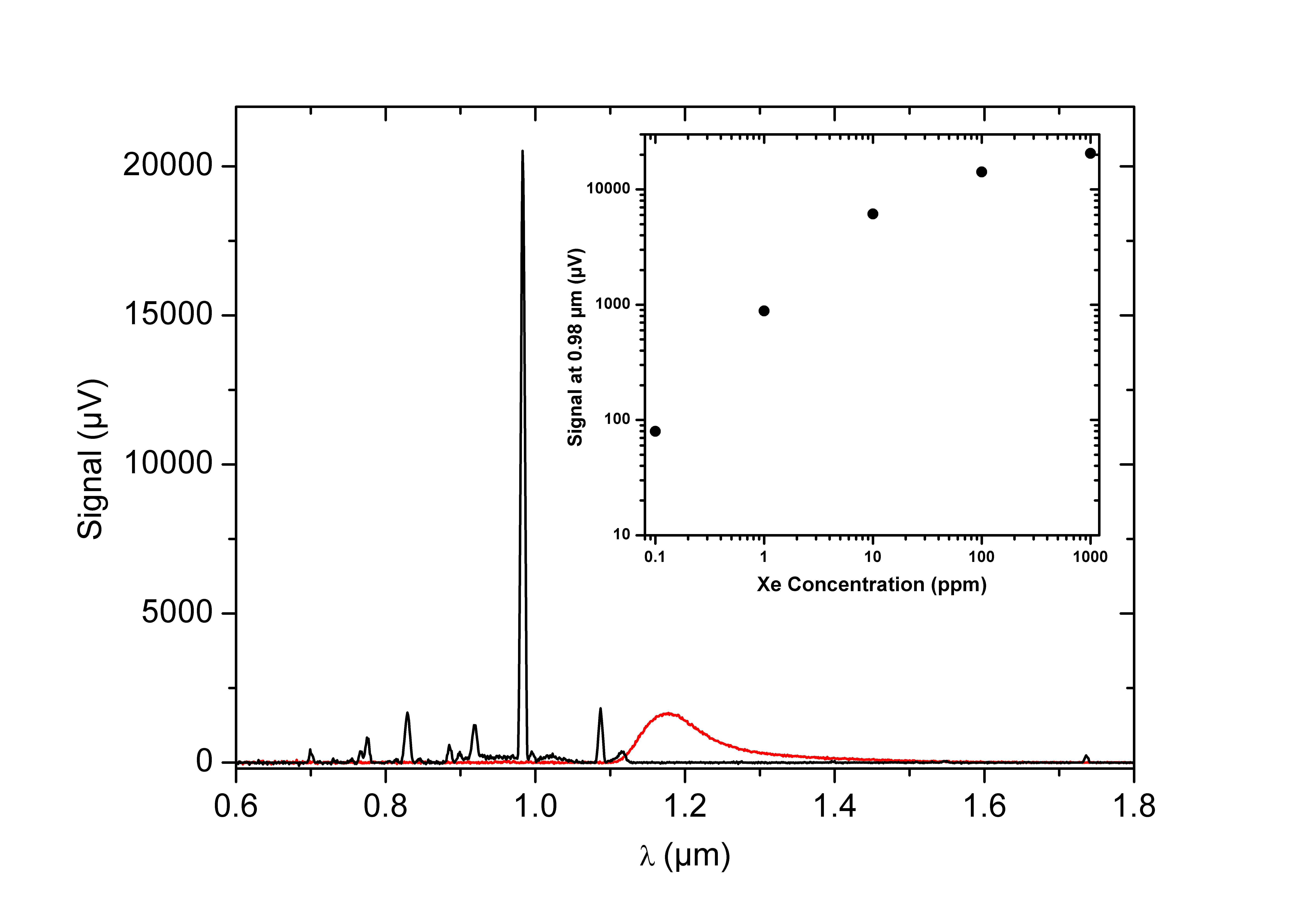}
 \caption{\textit{The electron-beam induced emission spectrum of 1000\,ppm xenon in gaseous argon (black line, T=100\,K, p=1255\,mbar) is shown as well as the emission spectrum of 10\,ppm xenon in liquid argon (red line). In the gas phase there is a strong energy transfer from argon to xenon visible in the most prominent XeI line (6s-6p) at 0.98\,$\mu$m. The dependence of the intensity of this line on the xenon concentration is shown in the inset.}}
 \label{fig:IR_Emission_GAr1000ppmXe}
\end{figure}

So far we have no clear assignment for the origin of the strong IR emission which has been found. In the gas phase the broad and asymmetric emission would clearly be an indication for an excimer-type transition of a bound molecule to a repulsive lower state. A reliable assignment will require detailed studies including pulsed excitation and time resolved spectroscopy as well as correlated spectroscopy in the VUV. A systematic variation of the xenon admixtures in small linear steps will show whether the energy transfer occurs as a two- or three-body collision and thereby whether ArXe$^{**}$ or Xe$_{2}^{**}$ is formed. Absorption studies may also help in that context. Energy transfer processes from the lighter to the heavier noble gases are of course well known from the gas phase \cite{Efthimiopoulos}. There is, for instance, a whole class of near-infrared lasers using dense noble gas mixtures as laser media, including Ar-Xe mixtures. They have been studied in the context of excimer physics as well as particle-
beam \cite{Ulrich_Nuc_He_Ar_Laser,Ulrich_ion_beam_Laser,Ulrich_e_beam_laser,Skrobol_e_beam_laser} and so called nuclear-pumped gas lasers \cite{Russen_Nuc_Laser}. In both cases a 1.73\,$\mu$m XeI-line is a strong and easy to operate laser line \cite{Ulrich_e_beam_laser,Skrobol_e_beam_laser}. To demonstrate the dramatic change in the emission spectrum between the gas and liquid phases we show in Fig.\,\ref{fig:IR_Emission_GAr1000ppmXe} a spectrum of the cold gas 
mixture of argon with 1000\,ppm xenon just before condensation.  In a series of emission spectra which were recorded in the gas-phase analogue to Fig.\,\ref{fig:3D_IR_Emission_LArXe} it turned out that the intensity of the most prominent XeI line (Racah notation: 6s\,[1\textonehalf]$_{2}$ - 6p\,[\textonehalf]$_{1}$) \cite{Sventitskii} at 0.98\,$\mu$m increases with xenon concentration (see inset in Fig.\,\ref{fig:IR_Emission_GAr1000ppmXe}) compared to the measurements in the liquid phase. The strong energy transfer from argon to a 6s-6p line in xenon had been observed before \cite{Italiener_IR_2,Italiener_IR_3}. {Whether} this line is {responsible} for the 1.18\,$\mu$m emission (observed in the liquid phase) and just shifted to this wavelength position{, or completely new} energy transfer mechanisms have to be invoked to explain the 1.18\,$\mu$m band{,} will have to be studied in the future. An emission feature in dense gaseous Ar-Xe mixtures found by Borghesani and 
coworkers \cite{Italiener_IR_2,Italiener_IR_3} is very 
similar in shape and with 1.3\,$\mu$m peak wavelength close to the emission at 1.18\,$\mu m$ found here. 
However, a simple density 
related effect shifting the emission to shorter wavelengths seems unlikely since the emission is redshifted with increasing pressure \cite{Italiener_IR_2,Italiener_IR_3}. {The 1.3\,$\mu m$ emission was observed in pure xenon as well as in argon with a 10\,\% xenon admixture and was attributed to Xe$_{2}^{**}$ by the authors of refs. \cite{Italiener_IR_2,Italiener_IR_3}.}

\section{Potential Applications}

The discovery of a second region (IR in addition to the VUV) with strong particle beam induced light emission from liquid noble gases may be of great importance for the particle detector development mentioned above. This is due to the fact that it is normally necessary to identify the particles which induce the scintillation light on an event by event basis. Typically two parameters have to be recorded which depend on the type of the incident particle to identify the projectiles. Measurements and a discussion of this issue concerning liquid noble gases can, for example, be found in one of our previous publications \cite{Hofmann}. Whether a comparison of the intensities or the time structure of the light emission between the VUV and the IR will be helpful in that context will be studied in the future. In this respect it is interesting to note that photomultipliers with cathodes sensitive at 1.18\,$\mu$m wavelength are available so that a scintillation detector with fast and sensitive detectors for both a VUV 
and an IR signal could be realized. Wavelength dependent quenching factors for other projectiles than electrons will be determined.

For the emission in the VUV we would make the following predictions: the Xe$_{2}$* VUV band will rise \cite{Heindl_2} and the Ar$_{2}$* emission in the VUV will decrease in intensity in liquid Ar-Xe mixtures in analogy to gas phase mixtures \cite{Efthimiopoulos}. A series of VUV emission spectra for various Ar-Xe gas mixtures is, e.g., shown in Fig. 7 of ref. \cite{Efthimiopoulos}. Wavelength integrated and time resolved measurements of the emission from xenon-doped liquid argon in the VUV can, e.g., be found in ref. \cite{Pollmann}.

\section*{Acknowledgements}
This experiment has been supported financially by the German Research Foundation (DFG) via the Excellence Cluster Universe and the Maier-Leibnitz-Laboratorium M\"unchen. The authors thank Mr. Rapp from the institute E10 at the Physik-Department for efficient support in electronic issues.


\begin{thebibliography}{99}
\bibliographystyle{unsrt}

  \bibitem{Heindl_1}
        T.\,Heindl et al.,
        EPL \textbf {91}, 62002 (2010)
        
  \bibitem{Excimer_Buch}
        Ch. K. Rhodes \and C.A. Brau, Excimer Lasers, Springer-Verlag, Berlin (1984)
        
  \bibitem{Heindl_2}
        T.\,Heindl et al.,
        JINST \textbf {6}, P02011 (2011)
        
  \bibitem{Hofmann}
        M. Hofmann  et al.,
        Eur. Phys. J. C \textbf{73}, 2618 (2013)
        
  \bibitem{Neumeier}
        A. Neumeier et al.,
        Eur. Phys. J. C, \textbf{72}, 2190 (2012)
  
  \bibitem{GERDA}
        M. Agostini  et al.,
        J. Phys.: Conf. Ser. \textbf{375}, 042009 (2012)
  
  \bibitem{EXO}
        M. Auger  et al.,
        JINST \textbf{7}, P05010 (2012)
  
  \bibitem{WARP}
        R. Acciarri  et al.,
        J. Phys.: Conf. Ser. \textbf{203}, 012006 (2010)
  
  \bibitem{ArDM_1}
        A. Rubbia  et al.,
        J. Phys.: Conf. Ser. \textbf{39}, 129 (2006)
  
  \bibitem{ArDM_2}
        A. Marchionni  et al.,
        J. Phys.: Conf. Ser. \textbf{308}, 012006 (2012)
  
  \bibitem{DarkSide}
        A. Wright (DarkSide Collaboration),\\
        arXiv:1109.2979v1[physics.ins-det] (2011)
  
  \bibitem{XENON100}
        E. Aprile  et al.,
        Astropart. Phys. \textbf{35}, 9 (2012)
  
  \bibitem{LUX}
        D. S. Akerib  et al.,
        Nucl. Instrum. Methods A \textbf{704}, 111 (2013)
  
  \bibitem{XENON1T}
        E. Aprile (XENON1T collaboration),\\ 
        arXiv:1206.6288 [astro-ph.IM](2013) 
  
  \bibitem{MiniCLEAN}
        A. Hime (MiniCLEAN Collaboration),\\
        arXiv:1110.1005v1[physics.ins-det] (2011)
        
  \bibitem{DEAP}
        M. G. Boulay, 
        J. Phys.: Conf. Ser. \textbf {375}, 012027 (2012)
  
  \bibitem{DARWIN_1}
        L. Baudis (DARWIN consortium),\\
        arXiv:1012.4764v1[astro-ph.IM] (2009)
  
  \bibitem{DARWIN_2}
        M. Schumann (DARWIN consortium)\\
        arXiv:1111.6251v1[astro-ph.IM] (2011)
  
  \bibitem{ICARUS}
        C. Rubbia  et al.,
        J. Instrum \textbf{6}, P07011 (2011)
  
  \bibitem{GLACIER}
        A. Rubbia  et al.,
        arXiv:hep-ph/0402110v1 (2004)
  
  \bibitem{LBNE}
        B. Yu  et al.,
        Phys. Proc. \textbf{37}, 1279 (2012)
  
  \bibitem{Italiener_IR_1}
        G. Bressi  et al.,
        Nucl. Instrum. Methods A \textbf{440}, 254 (2000)
  
  \bibitem{Russen_IR_1}
        A. Buzulutskov  et al.,
        EPL \textbf{94}, 52001 (2011)
  
  \bibitem{Russen_IR_2}
        A. Bondar  et al.,
        JINST \textbf{7}, P06014 (2012)
  
  \bibitem{Russen_IR_3}
        A. Bondar  et al.,
        JINST \textbf{7}, P06015 (2012)
  
  \bibitem{Wieser_Membran}
        J. Wieser  et al.,
        Rev. Sci. Instrum. \textbf{68}, 1360 (1997)
  
  \bibitem{Morozov_Effizienz}
        A. Morozov  et al.,
        J. Appl. Phys.\textbf{103}, 103301 (2008)
  
  \bibitem{Doke_Effizienz_1}
        T. Doke  et al., 
        Nucl. Instrum. Methods A \textbf{269}, 291 (1988)
  
  \bibitem{Doke_Effizienz_2}
        T. Doke  et al., Jpn. 
        J. Appl. Phys. \textbf{41}, 1538 (2002)
  
  \bibitem{Efthimiopoulos}
        T. Efthimiopoulos  et al., 
        J. Phys. D \textbf{30} (1997)
  
  \bibitem{Ulrich_Nuc_He_Ar_Laser}
        A. Ulrich  et al., 
        Appl. Phys. Lett. \textbf{42}, 782 (1983)
  
  \bibitem{Ulrich_ion_beam_Laser}
        A. Ulrich  et al., J. Appl. Phys. \textbf{63}, 2206 (1988)
  
  \bibitem{Ulrich_e_beam_laser}
        A. Ulrich  et al., 
        J. Appl. Phys. \textbf{86}, 3525 (1999)  
  
   \bibitem{Skrobol_e_beam_laser}
        C. Skrobol  et al.,
        Eur. Phys. J. D \textbf{54}, 103 (2009)
  
   \bibitem{Russen_Nuc_Laser}
        A. M. Voinov  et al.,
        Akademiia Nauk SSSR Doklady \textbf{245}, 80 (1983)
  
    \bibitem{Sventitskii}
        A.R. Striganov \and N.S. Sventitskii, 
        \textit{TABELS OF SPECTRAL LINES OF NEUTRAL AND IONIZED ATOMS}, Vol. 9, IFI/Plenum Data Corporation New York, Washington, p. 572 (1968)
  
    \bibitem{Italiener_IR_2}
        A. Borghesani  et al.,
        J. Chem. Phys. \textbf{115}, 13 (2001)
  
    \bibitem{Italiener_IR_3}
        A. Borghesani  et al.,
        Eur. Phys. J. D \textbf{35}, 299 (2005)
  
    \bibitem{Pollmann}
        P. Peiffer  et al.,
        JINST \textbf{3}, P08007 (2008)
\end{thebibliography}
\end{document}